\documentclass[reprint,aps,pre,superscriptaddress,amsmath,amssymb]{revtex4-2}

\usepackage[utf8]{inputenc}
\usepackage[T1]{fontenc}
\usepackage{graphicx}
\usepackage{dcolumn}
\usepackage{bm}
\usepackage{booktabs}
\usepackage{xcolor}
\usepackage{hyperref}

\begin{document}

\title{Two temperature scales in the Ising model on the $\{5,4\}$ hyperbolic
lattice with free boundaries: susceptibility peak and boundary-induced order}

\author{S. Jaroszewicz}
\affiliation{Comisión Nacional de Energía Atómica (CNEA), Buenos Aires, Argentina.}
\author{N. Mendez}
\email{nmendez@frh.utn.edu.ar}
\affiliation{Instituto Sábato, Universidad Nacional de San Martín, Bs. As., Argentina}

\author{Maria P. Beccar-Varela}
\affiliation{Department of Mathematical Sciences, UTEP. El Paso, United States}

\author{Maria Cristina Mariani}
\affiliation{Department of Mathematical Sciences, UTEP. El Paso, United States}

\date{\today}

\begin{abstract}
On the $\{5,4\}$ hyperbolic lattice, the outermost generation holds 73\% of the sites at all system sizes, making macroscopic averages strictly boundary-dependent. We study the ferromagnetic Ising model on this geometry with free boundaries by Monte Carlo simulation, demonstrating that conventional observables cease to identify a single critical scale. Instead, the finite lattices are organized by two distinct temperature scales. The susceptibility maximum identifies the lower transition scale at $T_{c2}=1.4782(15)J/k_{B}$ without extrapolation. However, the order-parameter distribution lacks a conventional fixed point at this scale. A distinct pseudocritical scale emerges instead at $T^{*}\simeq1.67(3)$ inside the boundary-sensitive intermediate phase, where Binder cumulant curves cross and effective exponents are compatible with mean-field criticality. Between $T_{c2}$ and $T^{*}$, the interior amplifies boundary fluctuations into induced order. Furthermore, because the volume grows exponentially with depth, finite-size scaling must be formulated in the generation index rather than the total number of sites. The lack of asymptotic power-law convergence in the scaling stretch confirms through an independent observable that $T^{*}$ is a finite-size crossover rather than a thermodynamic fixed point. Finally, by applying a coherent field to the outermost generation alone, we recover the upper thermodynamic transition at $T_{pt}=2.81(1)J/k_{B}$, demonstrating that it remains accessible under an appropriate boundary perturbation.
\end{abstract}

\maketitle

\section{Introduction}
\label{sec:intro}

Regular tessellations of the hyperbolic plane have attracted attention as
discretizations of anti-de Sitter space, in which a gravitational theory in
the bulk is dual to a conformal field theory on the
boundary~\cite{maldacena1998,Witten1998,Gubser1998,krioukov2010}, and, more
recently, as a setting in which negative curvature produces phases with no
Euclidean counterpart~\cite{Wang2025}. The geometric feature behind both
developments is that the boundary of a hyperbolic lattice is extensive. In a
Euclidean lattice in $d$ dimensions the number of surface sites grows as
$N^{(d-1)/d}$ and the boundary is a vanishing fraction of the system; on a
hyperbolic tessellation the number of sites grows exponentially with the
number of generations and the boundary fraction tends to a nonzero constant.
For the $\{5,4\}$ tessellation studied here that constant is
$1-1/(2+\sqrt{3})=0.732$, and the fraction is already $73\%$ at the smallest
size we simulate. Every macroscopic average is therefore a boundary average,
at every size, and the finite-size reasoning on which much of the standard
numerical toolkit rests---that surface corrections are subleading---is
unavailable.

The thermodynamics of the hyperbolic Ising model is well characterized. Shima
and Sakaniwa~\cite{shima2006}, Kr\v{c}m\'ar \textit{et al.}~\cite{Krcmar2008}
and Breuckmann \textit{et al.}~\cite{Breuckmann2020} established the phase
diagram and the static exponents, finding in the bulk the mean-field behavior
that the exponential growth of the number of neighbors makes natural. More
recent work has turned to the boundary. Asaduzzaman \textit{et
al.}~\cite{Asaduzzaman2022} found power-law boundary--boundary correlations
on $\{3,7\}$ tessellations on both sides of the bulk transition; Okunishi and
Nishino~\cite{Okunishi2024} showed by the corner transfer matrix
renormalization group (CTMRG) that bulk correlations decay exponentially while
boundary correlations retain power-law behavior; and Wang, Nussinov and
Ortiz~\cite{Wang2025} identified, under open boundary conditions, an
intermediate ``boundary-sensitive'' phase in which bulk order is induced by
the boundary rather than by spontaneous symmetry breaking. For the $\{5,4\}$
vertex lattice the two transitions bounding that phase are at $T_{c2}=1.479$
and $T_{\mathrm{pt}}=2.799$ in units of
$J/k_{\mathrm{B}}$~\cite{Wang2025,Krcmar2008} ($K_{c2}$ and $K_{c1}$ in the
notation of Ref.~\cite{Wang2025}), both obtained by CTMRG from observables at
the center of the system: the upper one under a fixed or perturbed boundary,
the lower one under an open boundary with the $\mathbb{Z}_2$ symmetry
enforced exactly.

The primary question we address is how to locate these transitions in a Monte
Carlo simulation with free boundaries, where every observable is an average
over a region that is mostly boundary. We measure the susceptibility in four
nested regions of very different boundary content---the outermost
generation, the whole system, the interior and the deep core---and require
that they agree. Individually they do not: the peak positions of the bulk
and of the core descend with system size while that of the boundary rises.
The total system is the exception. Its susceptibility maximum does not move
over six generations, $L=4$ to $9$, across which $N$ grows by a factor of
$739$, and a fit to a constant gives $T_{c2}=1.4782(15)\,J/k_{\mathrm{B}}$,
within $0.05\%$ of the CTMRG value, with no extrapolation. 

The finite lattices, however, are organized by a second temperature
scale as well. At $T_{c2}$ the distribution of the order parameter has no
fixed point: the magnetization vanishes only as $N^{-0.045}$, with an
exponent that is still decreasing, and the Binder cumulant increases with
size toward its ordered-phase value. A finite-size pseudocritical scale emerges at
$T^{\ast}\simeq1.67$, inside the boundary-sensitive phase, where the Binder
curves of the six sizes cross and where the effective exponents are compatible with the mean-field values $1/4$ and $1/2$. Both scales follow from the phase diagram of Ref.~\cite{Wang2025} once the
boundary is allowed to fluctuate: between $T_{c2}$ and $T^{\ast}$ the finite
lattices carry the induced order that characterizes the Eggarter phase of the
Cayley tree~\cite{Eggarter1974} and its hyperbolic
counterpart~\cite{Wang2025}, and $T^{\ast}$ is the temperature scale at which that
induced order gives way to the paramagnet in a lattice of finite depth. No region shows any structure at
$T_{\mathrm{pt}}$, as expected for a free boundary that carries three
quarters of the spins.

A secondary question is whether the upper transition of the boundary-sensitive
phase, which no region average detects, is accessible at all to a free-boundary simulation.
It is, once the boundary is probed coherently: the response of
each generation to a field applied on the outermost one locates
$T_{\mathrm{pt}}$ to better than a percent.

The paper is organized as follows. Section~\ref{sec:model} describes the
lattice, the Hamiltonian and the simulation protocol.
Section~\ref{sec:thermodynamics} outlines the scaling properties associated with the geometry, locates the two temperature scales $T_{c2}$ and $T^{\ast}$, and shows how the upper transition can be recovered under a perturbed boundary. Section~\ref{sec:conclusions} summarizes our findings.

\section{Model, geometry and simulation protocol}
\label{sec:model}

\subsection{Hyperbolic lattice construction}

We consider the ferromagnetic Ising model on a discretized hyperbolic plane.
The geometry is a regular tessellation of the Poincar\'e disk denoted by the
Schl\"afli symbol $\{p,q\}$, where $p$ is the number of sides of each polygon
and $q$ the number of polygons meeting at each vertex; the tessellation is
hyperbolic when $(p-2)(q-2)>4$. We work with $\{5,4\}$, for which
$(p-2)(q-2)=6$. Spins are placed on the \emph{vertices} of the tessellation,
so that every interior site has coordination number $z_i=q=4$. This choice
combines a coordination number equal to that of the square lattice---which
makes the comparison with Euclidean results direct---with a pronounced negative
curvature, and it is the lattice for which transition temperatures are
available in the literature~\cite{Krcmar2008,Wang2025}.

We state the placement of the spins explicitly because the same tessellation
supports a second, inequivalent spin graph. Assigning spins instead to the
pentagonal \emph{cells} of $\{5,4\}$, with two cells coupled when they share an
edge, produces a graph that is $5$-regular in the interior---the dual $\{4,5\}$
lattice---whose transition temperatures are different. Only the vertex graph is
directly comparable with the corner transfer matrix renormalization group
(CTMRG) results quoted below.

The lattice is generated with the \texttt{hypertiling}
library~\cite{Schrauth2024} (version 1.4.1), exploiting the
duality between the two tessellations: the cells of $\{4,5\}$, with adjacency
defined by shared edges, are in one-to-one correspondence with the vertices of
$\{5,4\}$ together with their bonds, since each quadrilateral of $\{4,5\}$ has
four edges. We therefore generate the $\{4,5\}$ tessellation and read off its
cell-adjacency list, which is returned already symmetric. 

The construction is verified at every $L$ by four assertions. (i) The adjacency
list is symmetric, $j\in\mathcal{N}(i)\Leftrightarrow i\in\mathcal{N}(j)$.
(ii) The degree distribution is exactly $\{2,3,4\}$, with $z_i=4$ in the
interior. (iii) The number of sites with $z_i=4$ equals $N(L-1)$ exactly, so
that the interior of a lattice with $L$ generations reproduces the full lattice
with $L-1$ generations. (iv) The set $\{i: z_i<4\}$ coincides exactly with the
outermost generation; the two possible definitions of the boundary are
therefore equivalent, which is a property of this tessellation rather than a
convention, and we adopt $z_i<q$ throughout.

\begin{figure*}[t]
    \centering
    \begin{minipage}{0.45\textwidth}
        \centering
        \includegraphics[width=\linewidth]{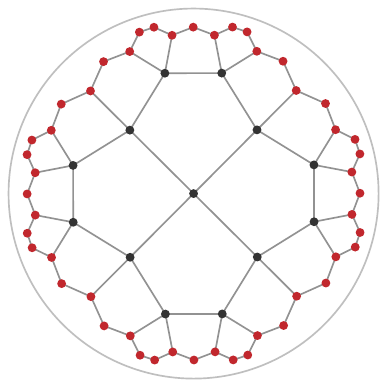}
        \textbf{(a) $L=3$ ($N=61$)}
    \end{minipage}\hfill
    \begin{minipage}{0.45\textwidth}
        \centering
        \includegraphics[width=\linewidth]{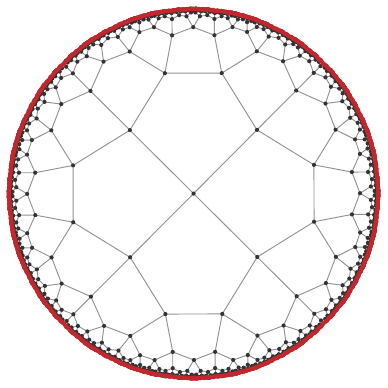}
        \textbf{(b) $L=6$ ($N=3421$)}
    \end{minipage}
    \caption{\textbf{The $\{5,4\}$ hyperbolic lattice on the Poincar\'e disk.}
Spins occupy the vertices of the tessellation, drawn here as the sites at which
$q=4$ pentagonal faces meet, so that interior sites (dark) have coordination
$z_i=4$; boundary sites (red) belong to the outermost generation and have
$z_i<4$. (a)~$L=3$ ($N=61$), with the pentagonal faces shaded to make the local
connectivity explicit. (b)~$L=6$ ($N=3421$); the exponential growth of the
number of sites with each generation is apparent, and the boundary already
occupies $73.3\%$ of the lattice, close to the asymptotic value of
Eq.~\eqref{eq:boundary_fraction}, $73.21\%$.}
    \label{fig:lattices}
\end{figure*}

Sites are added in concentric generations $k=1,\dots,L$ starting from a central
site. The number of sites added in generation $k$ obeys the recursion
$d_{k+1}=4d_{k}-d_{k-1}$, so that $d_{k+1}/d_{k}\to 2+\sqrt{3}\simeq 3.7321$
and both $N$ and $N_{\partial}$ grow exponentially with $L$ at the same rate.
Consequently the boundary fraction does not vanish in the thermodynamic limit
but saturates at
\begin{equation}
\lim_{L\to\infty}\frac{N_{\partial}}{N}=1-\frac{1}{2+\sqrt{3}}=0.73205 .
\label{eq:boundary_fraction}
\end{equation}
Table~\ref{tab:boundary_ratio} lists the resulting site counts, and
Fig.~\ref{fig:lattices} shows the lattice for $L=3$ and $L=6$.

\begin{table}[t]
\centering
\caption{Geometry of the $\{5,4\}$ vertex lattice as a function of the number
of generations $L$: total number of sites $N$, number of boundary sites
$N_{\partial}$, boundary fraction, and the sizes of the bulk and core regions
defined in the text. All entries are exact counts obtained from the generated
adjacency lists. The boundary fraction converges to the value given in
Eq.~\eqref{eq:boundary_fraction}.}
\label{tab:boundary_ratio}
\begin{tabular}{cccccc}
\toprule
$L$ & $N$ & $N_{\partial}$ & $N_{\partial}/N$ (\%) & $N_{\mathrm{bulk}}$ & $N_{\mathrm{core}}$\\
\midrule
4 & 241     & 180     & 74.69 & 61     & 13\\
5 & 913     & 672     & 73.60 & 241    & 61\\
6 & 3\,421  & 2\,508  & 73.31 & 913    & 241\\
7 & 12\,781 & 9\,360  & 73.23 & 3\,421 & 913\\
8 & 47\,713 & 34\,932 & 73.21 & 12\,781& 3\,421\\
9 & 178\,081& 130\,368& 73.21 & 47\,713& 12\,781\\
\bottomrule
\end{tabular}
\end{table}

Table~\ref{tab:boundary_ratio} makes explicit the geometric fact that governs
everything that follows. The boundary fraction is already within $0.15\%$ of
its asymptotic value at $L=6$ and is indistinguishable from it at $L=8$ and
$L=9$. In Euclidean lattices $N_{\partial}/N\sim N^{-1/d}\to 0$, and enlarging
the system eventually produces a regime in which bulk behavior dominates. Here
no such regime exists: at every accessible size roughly three out of four spins
lie on the outermost generation and have reduced coordination. Increasing $L$
rescales $N$ but leaves the relative weight of the interior unchanged. This is
not a finite-size limitation to be removed by simulating larger systems; it is
a property of the geometry, and it determines which transitions are accessible
under open boundary conditions, as we show in Sec.~\ref{sec:thermodynamics}.

Accordingly, all observables in this work are resolved into four nested spatial
regions. The \emph{boundary} is the outermost generation, equivalently the set
$\{z_i<4\}$, of size $N_{\partial}$. The \emph{bulk} is its complement, the
$N_{\mathrm{bulk}}=N(L-1)$ interior sites with $z_i=4$. The \emph{core} is
obtained by removing the two outermost generations,
$N_{\mathrm{core}}=N(L-2)$; since no core site is adjacent to a boundary site,
it is the most stringent probe of interior behavior available at a given $L$.
The \emph{total} region comprises all $N$ sites. Reporting all four is
essential in this geometry: as shown in Sec.~\ref{sec:thermodynamics}, the four
regions carry finite-size biases of opposite sign, and their agreement is a
stronger statement than any one of them taken alone.

\subsection{Hamiltonian and dynamics}

The system is governed by the ferromagnetic Ising Hamiltonian with open
boundary conditions,
\begin{equation}
H = -J \sum_{\langle i,j\rangle} \sigma_i \sigma_j,
\label{eq:Ising_H}
\end{equation}
where $\sigma_i=\pm 1$ is the spin at site $i$, $J>0$ is the nearest-neighbor
coupling, and the sum runs over the bonds of the lattice defined in
Sec.~\ref{sec:model}. Interior sites contribute four terms each and boundary
sites two or three, according to the degree distribution reported above.

Thermal fluctuations are sampled by single-spin-flip Monte Carlo dynamics in
contact with a heat bath at temperature $T$. We supplement local Metropolis dynamics with
Wolff cluster updates~\cite{Wolff1989}, in which bonds are activated with probability
$p_{\mathrm{add}}=1-e^{-2\beta J}$. A hybrid Wolff--Metropolis
dynamics is used exclusively for equilibrium averages, avoiding the critical slowing down and the exponentially long time to tunnel between ordered states in finite systems with unbroken $\mathbb{Z}_2$ symmetry.

\subsection{Computational protocol}
\label{sec:computational_protocol}

Simulations are initialized from a fully ordered configuration $\sigma_i=+1$
and equilibrated with $N_{\mathrm{eq}}=2\times10^{4}$ hybrid steps, each
consisting of one Wolff cluster update followed by one Metropolis sweep of $N$
attempted flips. Measurements are then accumulated over
$N_{\mathrm{mc}}=4\times10^{4}$ further hybrid steps, recording one measurement
per step. A single chain is run for each pair $(L,T)$, and statistical errors
are obtained from a jackknife analysis over $50$ blocks of that chain.

The quantity that controls the statistical quality of a run is the number of
independent samples it contains, not the number of steps, and we therefore
measure the integrated autocorrelation time $\tau_{\mathrm{int}}$ directly from
the time series rather than assume it. Which observable is used matters here.
The Wolff bond probability is $p_{\mathrm{add}}=1-e^{-2J/k_{\mathrm{B}}T}$,
which equals $0.742$ at $T_{c2}$, $0.698$ at $T^{\ast}$ and $0.511$ at
$T_{\mathrm{pt}}$. The bond percolation threshold of a $z=4$ lattice lies
between $1/3$, its value on the Bethe lattice of the same coordination, and
$1/2$, its value on the square lattice, and negative curvature places the
$\{5,4\}$ tessellation nearer the former. The cluster therefore spans the
system at every temperature of interest, the sign of $m$ reverses on
essentially every update, and $\tau_{\mathrm{int}}$ of the signed magnetization
measures the rate of that global $\mathbb{Z}_2$ flip and nothing else. We
consequently quote $\tau_{\mathrm{int}}$ for $|m_R|$, $m_R^2$, $m_R^4$ and the
energy, which are the quantities entering Eqs.~\eqref{eq:chi_region}
and~\eqref{eq:Binder_def}. It is obtained with the automatic windowing
procedure of Madras and Sokal~\cite{MadrasSokal1988},
$W=\min\{W: W\ge c\,\tau_{\mathrm{int}}(W)\}$ with $c=6$, and its uncertainty
from $\delta\tau_{\mathrm{int}}=\tau_{\mathrm{int}}[2(2W+1)/N_{\mathrm{mc}}]^{1/2}$.
Table~\ref{tab:tau} collects the result at $L=9$, the largest size and the
one for which the demand on the protocol is greatest. The longest time we find
is $\tau_{\mathrm{int}}=20.4(9)$ hybrid steps for $|m|$ at $T=1.86$, so that
$N_{\mathrm{mc}}=4\times10^{4}$ contains at least $10^{3}$ autocorrelation
times and each jackknife block is longer than $98\,\tau_{\mathrm{int}}$, at
every temperature simulated. The table also confirms the argument of the
preceding paragraph: below the transition region $\tau_{\mathrm{int}}$ of the
signed magnetization is less than half that of $|m|$, exactly as expected when
a percolating cluster reverses the sign on most updates, so that the signed
quantity would have understated the correlation time by a factor of two.
One hybrid step is one cluster update followed by one local sweep, so these
times are not directly comparable with values quoted in units of local sweeps.

Three independent checks bound the error bars themselves, none of which relies
on the estimate of $\tau_{\mathrm{int}}$. First, the six determinations of
$T_\chi^{\mathrm{tot}}$ in Table~\ref{tab:peaks} are statistically
independent runs: their scatter about the weighted mean, $0.0028\,J/k_{\mathrm{B}}$,
coincides with the mean quoted uncertainty, $0.0028\,J/k_{\mathrm{B}}$, and the
one-parameter fit of Eq.~\eqref{eq:Tc2} gives $\chi^2/\nu=1.3$ for $\nu=5$.
Error bars underestimated by a factor $1.5$ would raise this to $\chi^2/\nu=2.9$
with $p=0.013$, and the data bound that factor to $2.36$ at $95\%$ confidence.
Second, the collapse statistic of Sec.~\ref{sec:scaling_variable} attains
$S=1.10(9)$; a fit whose error bars were misestimated by any substantial factor
would not land on unity. Third, the procedure has been validated end to end: cutting long chains at $L=9$
into segments of $4\times10^{4}$ steps and analyzing each exactly as a
production run, the quoted jackknife error and the actual scatter of
$U_4^{\mathrm{tot}}$ across segments agree, with a ratio between $0.78$ and
$1.44$ over the four temperatures studied and $\chi^{2}=4.0$ for four degrees
of freedom against a ratio of unity ($p=0.41$)..

We simulate $L=4,\dots,9$, that is $N=241$ to $N=178\,081$, on a grid of $78$
temperatures spanning $0.90\le T\le 3.60$ in units of $J/k_{\mathrm{B}}$. The
grid is refined to $\Delta T=0.01$ over $[1.32,1.62]$, which brackets $T_{c2}$,
to $\Delta T=0.02$ over $[1.62,1.90]$, and to $\Delta T=0.05$ over
$[2.65,3.00]$, which brackets $T_{\mathrm{pt}}$. Covering both reference
temperatures in a single sweep is necessary here: with $73\%$ of the sites on
the boundary, a window chosen around one of them gives no information about
whether the other produces any structure at all.

For each region $R\in\{\mathrm{bulk},\partial,\mathrm{total},\mathrm{core}\}$
we record the instantaneous magnetization per site restricted to that region,
$m_R = N_R^{-1}\sum_{i\in R}\sigma_i$, and compute the susceptibility
\begin{equation}
\chi_R(T,L) = \frac{N_R}{k_{\mathrm{B}} T}
\Bigl( \langle m_R^2 \rangle - \langle |m_R| \rangle^2 \Bigr).
\label{eq:chi_region}
\end{equation}
The absolute value in the second term is required, and is the standard
convention for systems with a discrete $\mathbb{Z}_2$ symmetry~\cite{Binder1981}:
in a finite system at zero field the two ordered states are sampled with equal
weight, so $\langle m_R\rangle\to 0$ and the subtracted term would vanish
identically, leaving $\chi_R$ proportional to $\langle m_R^2\rangle$ and
therefore not an estimator of the susceptibility. The fourth-order Binder
cumulant, defined in Sec.~\ref{sec:binder_analysis}, involves only even moments
and is unaffected by this choice.

\section{Thermodynamics and Temperature Scales}
\label{sec:thermodynamics}

Two features of the geometry make mapping the phase transition less routine
than in a Euclidean lattice. First, there is no unique linear length scale: the number of sites grows exponentially with the
number of generations. As outlined in Eq.~\eqref{eq:boundary_fraction}, $N(L) \sim \lambda^L$ with $\lambda = 2+\sqrt{3}$. Using $N$ as the finite-size scaling variable is profoundly different from ordinary Euclidean scaling where $N = L^d$, but it serves as the natural extensive measure of the system size here. Second, the boundary is a finite fraction of the system at every size, so no observable averaged over a macroscopic region is free of boundary contributions.

Our strategy is to measure the same quantity in four
regions of very different boundary content and to require that the answers agree. Where they do, the
agreement is evidence; where they do not, the disagreement is a measurement of
how strongly the estimator is contaminated by the boundary. 

The simulation scheme is validated against the exact
solution of the $\{5,4\}$ lattice on small clusters and against published
results on the square lattice in Appendix~\ref{app:validation}.

\subsection{The susceptibility peak identifies the lower transition scale $T_{c2}$}
\label{sec:peak}

For each size and each region we locate the maximum of $\chi_R(T,L)$,
Eq.~\eqref{eq:chi_region}, by a weighted parabolic fit. We fit all contiguous points with
$\chi\ge0.90\,\chi_{\max}$, which adapts the fitting range to the width of the
peak, and estimate the uncertainty by resampling $\chi$ within its jackknife
error and discarding realizations whose vertex falls outside the fitted range.
Varying the threshold between $0.80$ and $0.90$ moves the resulting temperature by $0.0010$, which we carry as a systematic uncertainty.

\begin{figure*}[t]
\centering
\includegraphics[width=0.9\textwidth]{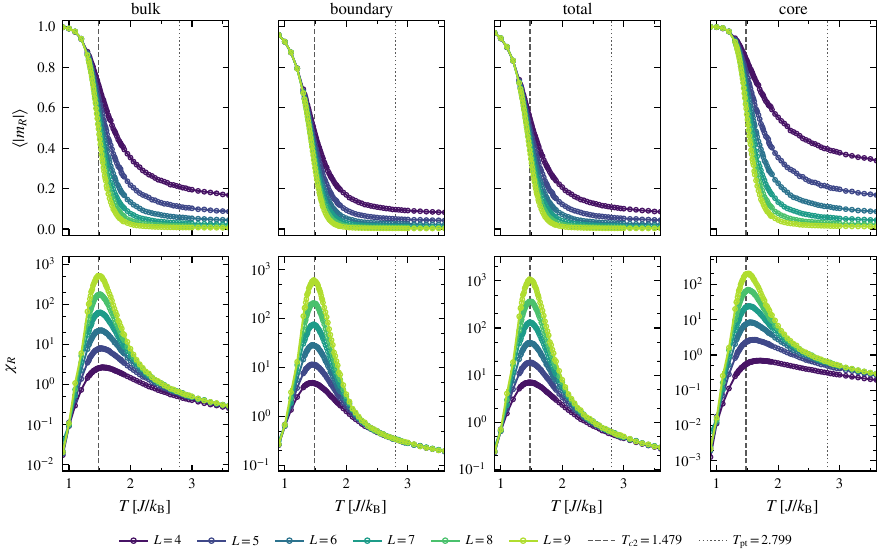}
\caption{\textbf{Magnetization and susceptibility of the $\{5,4\}$ hyperbolic
Ising model.} $\langle|m_R|\rangle$ (top) and $\chi_R$ (bottom) for
$L=4,\dots,9$ in the four regions of Sec.~\ref{sec:model}. Since
$\chi_R\propto N_R$ and $N_R$ differs by up to three orders of magnitude
between regions, the vertical scale of each susceptibility panel is set
independently; the magnetization panels share a common scale. Vertical lines mark
the published transition temperatures $T_{c2}=1.479$~\cite{Wang2025} and
$T_{\mathrm{pt}}=2.799$~\cite{Krcmar2008}. The susceptibility maximum of the
total system sits on $T_{c2}$ at every size; the maxima of the other three
regions approach it from opposite sides (Sec.~\ref{sec:peaks_regions}). No
region shows any structure at $T_{\mathrm{pt}}$
(Sec.~\ref{sec:upper_transition}).}
\label{fig:fss_regions}
\end{figure*}

\begin{figure*}[t]
\centering
\includegraphics[width=0.9\textwidth]{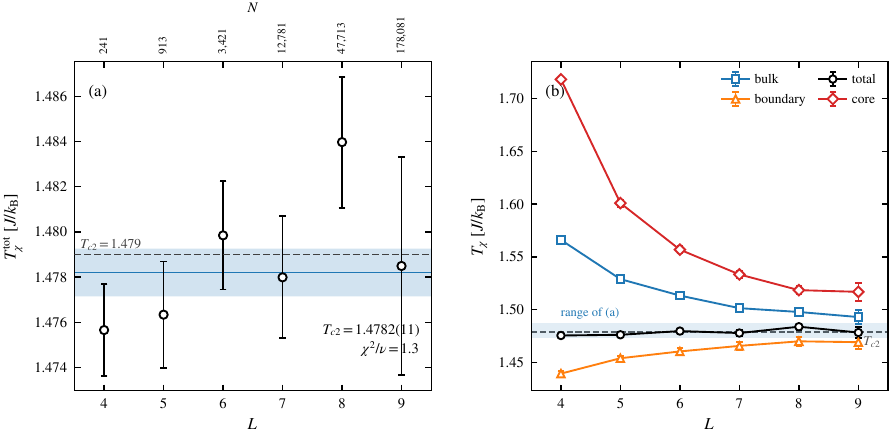}
\caption{\textbf{The susceptibility peak requires no extrapolation.}
(a) Position of the maximum of $\chi_{\mathrm{total}}$ against $L$, with the
constant fit of Eq.~\eqref{eq:Tc2} (band) and the published value
$T_{c2}=1.479$ (dashed line). (b) The same for the four regions: bulk and core
descend, boundary rises, total is flat, and all four extrapolate to the same
place (Table~\ref{tab:peaks}). The abscissa is the generation index; note that
$N$ spans more than two and a half decades across the panel.}
\label{fig:peak_flat}
\end{figure*}

The magnetization and susceptibility curves of the four regions are shown in
Fig.~\ref{fig:fss_regions}, and the peak positions are collected in
Table~\ref{tab:peaks} and displayed in Fig.~\ref{fig:peak_flat}. The result
for the total system is that they do not move:
\begin{equation}
\begin{split}
T_{\chi}^{\mathrm{tot}}(L) = {}& 1.4757,\ 1.4763,\ 1.4798,\\
& 1.4780,\ 1.4840,\ 1.4785
\end{split}
\nonumber
\end{equation}
for $L=4$ to $9$, while $N$ increases by a factor of $739$. A fit to a constant
gives
\begin{equation}
\begin{split}
T_{c2} &= 1.4782 \pm 0.0011\,(\mathrm{stat}) \pm 0.0010\,(\mathrm{syst})\\
       &= 1.4782(15)\,J/k_{\mathrm{B}},
\end{split}
\label{eq:Tc2}
\end{equation}
with $\chi^2/\nu=1.3$ for $\nu=5$. The published CTMRG value for this lattice
is $T_{c2}=1/K_{c2}=1.479$~\cite{Wang2025}; Eq.~\eqref{eq:Tc2} deviates from it
by $0.05\%$, well within one standard deviation.

We stress what this determination does not involve. There is no extrapolation,
no functional form to select, no correction-to-scaling exponent to fit, and no
free parameter beyond the constant itself. The consistency of the error budget
is internal: with the quoted per-point uncertainties, a one-parameter constant
model describes six independent measurements with $\chi^2/\nu=1.3$.

\begin{table}[t]
\centering
\caption{Position of the susceptibility maximum, $T_{\chi}$, in units of
$J/k_{\mathrm{B}}$, for the four regions and six generations. Parenthesized
digits are uncertainties in the last quoted digits, obtained by resampling
$\chi$ within its jackknife error. The row $\chi^2/\nu$ gives the quality of a
fit of each column to a constant ($\nu=5$); only the total system is compatible
with an $L$-independent peak. The last row gives the $L\to\infty$ extrapolation
obtained with the model-selection protocol of Sec.~\ref{sec:peaks_regions},
including the systematic spread among admissible functional forms. The
reference value is $T_{c2}=1.479$~\cite{Wang2025}.}
\label{tab:peaks}
\begin{tabular}{lcccc}
\toprule
$L$ & bulk & boundary & total & core\\
\midrule
4 & 1.5661(23) & 1.4396(21) & 1.4757(20) & 1.7180(26)\\
5 & 1.5290(21) & 1.4542(25) & 1.4763(23) & 1.6009(33)\\
6 & 1.5134(25) & 1.4606(31) & 1.4798(25) & 1.5569(33)\\
7 & 1.5015(38) & 1.4658(32) & 1.4780(26) & 1.5334(34)\\
8 & 1.4979(27) & 1.4701(45) & 1.4840(30) & 1.5186(35)\\
9 & 1.4931(68) & 1.4694(63) & 1.4785(45) & 1.5169(85)\\
\midrule
$\chi^2/\nu$ (constant) & 101 & 17 & \textbf{1.3} & 650\\
$L\to\infty$ & 1.485(11) & 1.478(15) & 1.4841(38) & 1.507(9)\\
\bottomrule
\end{tabular}
\end{table}

\subsection{Four regions with finite-size biases of opposite sign}
\label{sec:peaks_regions}

The remaining columns of Table~\ref{tab:peaks} turn the flatness of the total
peak from a fortunate coincidence into a controlled result.

The four regions differ enormously in boundary content: the boundary region is
entirely composed of sites with $z_i<4$, the total system contains $73\%$ of
them, the bulk none, and the core is separated from the boundary by an entire
generation. Their susceptibility maxima behave accordingly. Bulk and core
approach $T_{c2}$ \emph{from above} and monotonically, from $1.5661$ to
$1.4931$ and from $1.7180$ to $1.5169$ respectively; the boundary approaches it
\emph{from below}, from $1.4396$ to $1.4694$, and has essentially stopped
moving by $L=8$; the total system does not move at all. Fitting each column to
a constant quantifies this: $\chi^2/\nu=1.3$ for the total against $17$, $101$
and $650$ for boundary, bulk and core.

That the four sequences drift in opposite directions and at very different
rates, and nevertheless converge on the same limit, is a stronger statement
than the agreement of any single estimator with the literature value. A
systematic effect common to all four would displace them together; an accident
would not reproduce itself four times from four different directions.

Making the convergence quantitative requires extrapolating three sequences that
do drift, and here the choice of functional form is itself a source of
uncertainty that has to be reported. The natural ansatz,
\begin{equation}
T_{\chi}(L) = T_{\chi}^{\infty} + a\,L^{-\omega},
\label{eq:Tc_scaling}
\end{equation}
is degenerate in the regime relevant here: as $\omega\to0$ one has
$aL^{-\omega}\simeq a(1-\omega\ln L)$, so the model collapses onto
$(T^{\infty}+a)-a\omega\ln L$, the fit quality becomes excellent and the
extrapolated intercept is the difference of two large and strongly
anticorrelated numbers. The same degeneracy affects $a\,e^{-cL}$ as $c\to0$.
Selecting the form with the smallest $\chi^2/\nu$ therefore selects the
degenerate one.

We instead fit $L^{-1}$, $L^{-2}$, $L^{-\omega}$ and $e^{-cL}$, and reject any
fit whose extrapolated value falls outside the range of the data, whose
correction term at the smallest $L$ exceeds fifty times that range, whose
exponent is pinned at zero, or whose extrapolation error exceeds the range
itself. Of the admissible forms we report the spread as a systematic
uncertainty in addition to the statistical error of each fit. For the total
system, $L^{-\omega}$ is rejected (extrapolation outside the data range) and
$e^{-cL}$ is rejected (error larger than the range), while $L^{-1}$ and
$L^{-2}$ give $1.4861(42)$ and $1.4822(23)$ with $\chi^2/\nu=0.64$ and $0.67$,
for a combined $1.4841(38)$.

The four extrapolations, listed in the last row of Table~\ref{tab:peaks}, are
$1.485(11)$, $1.478(15)$, $1.4841(38)$ and $1.507(9)$. They span $0.029$ in
$T$, or $2\%$, and bracket the published value. We take Eq.~\eqref{eq:Tc2}, the
constant fit to the total system, as our determination, since it is the one
column that requires no extrapolation at all.

\subsection{Finite-size scaling of the peak height and of the order parameter}
\label{sec:peak_height}

The position of the susceptibility maximum is only half of what
Fig.~\ref{fig:fss_regions} contains. Its height, and the magnetization at
$T_{c2}$, scale with system size in a way that bears directly on how the
transition should be read. We
quote local exponents between successive generations,
$\gamma_{\mathrm{eff}}=\Delta\ln\chi_R^{\max}/\Delta\ln N_R$ and
$\beta_{\mathrm{eff}}=-\Delta\ln\langle|m_R|\rangle/\Delta\ln N_R$, rather
than a single fitted power, because most of them drift. The reference values
are those of mean-field finite-size scaling with $N$ as the scaling
variable~\cite{BotetJullienPfeuty1982}: $\chi^{\max}\propto N^{1/2}$ and
$\langle|m|\rangle\propto N^{-1/4}$ at the critical point, and
$\langle|m|\rangle\propto N^{-1/2}$ in the paramagnetic phase, where the
magnetization is a sum of uncorrelated variables. The peak height is taken as
the largest measured value of $\chi_R$ with its jackknife error; the vertex of
the parabolic fit of Sec.~\ref{sec:peak} lies $2$--$4\%$ lower at every size
and gives the same exponents to within $0.02$. The magnetization at $T_{c2}$
is interpolated from the $\Delta T=0.01$ grid, with an uncertainty estimated
from the residuals of a local polynomial fit.

\begin{table*}[t]
\centering
\caption{Local finite-size exponents between successive generations, with
$N_R$ as the scaling variable: $\gamma_{\mathrm{eff}}$ from the height of the
susceptibility maximum and $\beta_{\mathrm{eff}}$ from the magnetization at
$T_{c2}=1.4782$. Mean-field finite-size scaling at a critical point gives
$1/2$ and $1/4$ respectively. The bulk column of $\gamma_{\mathrm{eff}}$ is
the only one compatible with a constant; a power law fitted to $L=5$--$9$
gives $0.783(4)$ with $\chi^{2}/\nu=1.1$. At the largest interval all four
regions agree.}
\label{tab:exponents}
\begin{tabular}{lcccccccc}
\toprule
& \multicolumn{4}{c}{$\gamma_{\mathrm{eff}}$ ($\chi_R^{\max}$)} &
  \multicolumn{4}{c}{$\beta_{\mathrm{eff}}$ ($\langle|m_R|\rangle$ at $T_{c2}$)}\\
\cmidrule(lr){2-5}\cmidrule(lr){6-9}
$L\to L+1$ & boundary & total & bulk & core & boundary & total & bulk & core\\
\midrule
$4\to5$ & 0.657(10) & 0.727(12) & 0.795(11) & 0.882(9)  & 0.084(4) & 0.073(4) & 0.065(3) & 0.052(3)\\
$5\to6$ & 0.695(16) & 0.720(13) & 0.768(11) & 0.826(13) & 0.067(3) & 0.063(3) & 0.060(3) & 0.053(3)\\
$6\to7$ & 0.735(21) & 0.768(18) & 0.780(13) & 0.795(16) & 0.057(3) & 0.055(3) & 0.054(3) & 0.051(3)\\
$7\to8$ & 0.766(20) & 0.770(21) & 0.793(18) & 0.805(19) & 0.047(3) & 0.047(3) & 0.046(3) & 0.045(3)\\
$8\to9$ & 0.824(24) & 0.829(24) & 0.801(24) & 0.812(20) & 0.044(3) & 0.043(3) & 0.043(3) & 0.043(3)\\
\bottomrule
\end{tabular}
\end{table*}

\begin{figure*}[t]
\centering
\includegraphics[width=\textwidth]{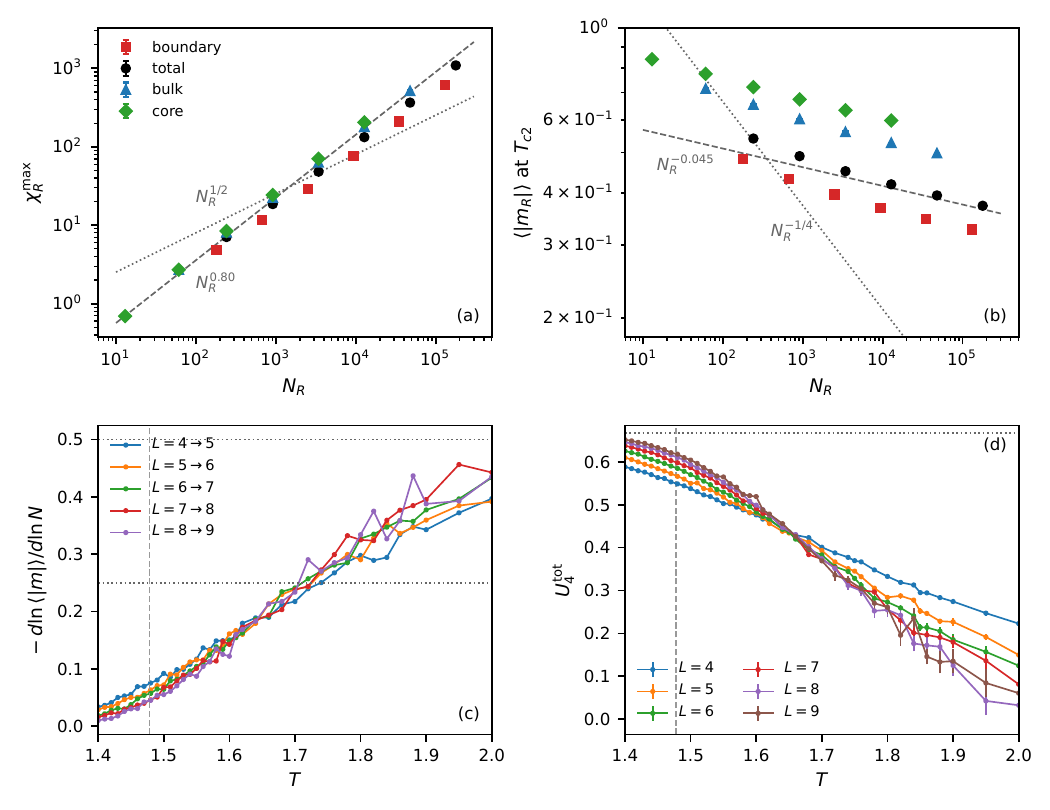}
\caption{Finite-size scaling of the order parameter. (a) Height of the
susceptibility maximum against the number of sites of the region; the dashed
line is $N_R^{0.80}$ and the dotted line the mean-field $N_R^{1/2}$.
(b) Magnetization at $T_{c2}$ against $N_R$; the dashed line is
$N_R^{-0.045}$ and the dotted line the mean-field $N_R^{-1/4}$.
(c) Local exponent of the total magnetization,
$-\Delta\ln\langle|m|\rangle/\Delta\ln N$, for the five size intervals as a
function of temperature; horizontal lines mark the mean-field critical value
$1/4$ and the paramagnetic value $1/2$, the vertical line marks $T_{c2}$. The
five curves cross at $T^{\ast}\simeq1.67$. (d) Binder cumulant of the total
magnetization for the six sizes over the same range; the curves cross in
$1.64\le T\le1.70$ and separate monotonically on both sides. At $T_{c2}$ the
cumulant increases with size toward $2/3$.}
\label{fig:peak_height_fss}
\end{figure*}

Table~\ref{tab:exponents} and Fig.~\ref{fig:peak_height_fss}(a) give the
result for the peak height. In the bulk the local exponent is stable,
$0.77$--$0.80$ over all five intervals, and a single power law fitted to
$L=5$--$9$ gives $\gamma_{\mathrm{eff}}=0.783(4)$ with $\chi^{2}/\nu=1.1$. The
other three regions drift, with the same opposite signs that govern the peak
positions in Table~\ref{tab:peaks}: the boundary rises from $0.66$ to $0.82$,
the total from $0.73$ to $0.83$, and the core falls from $0.88$ to $0.81$. At
the largest interval, $L=8\to9$, all four regions lie between $0.80$ and
$0.83$, within their uncertainties of a common value
$\gamma_{\mathrm{eff}}\simeq0.81$. The convergence of drifting sequences onto
one value from both sides makes the data consistent with an effective scaling $\chi^{\max}\propto N^{0.8}$ over the simulated range. The value is far from the mean-field $N^{1/2}$, and since
$\chi_R^{\max}/N_R$ is the variance of $|m_R|$ at the peak, it means that the
fluctuations of the order parameter at $T_{c2}$ decay only as $N^{-0.2}$.

The magnetization at $T_{c2}$, Fig.~\ref{fig:peak_height_fss}(b), completes
the picture. $\langle|m_R|\rangle$ decreases with a local exponent of
$0.04$--$0.08$ that itself decreases with size, and at the last interval
takes the same value, $0.043$--$0.044$, in all four regions. At $N=178\,081$
the magnetization of the total system is still $0.37$, and the Binder
cumulant at $T=1.48$ increases monotonically with size, from
$U_4^{\mathrm{tot}}=0.549(2)$ at $L=4$ to $0.618(1)$ at $L=9$, toward the
ordered-phase value $2/3$ [Fig.~\ref{fig:peak_height_fss}(d)]. None of these
quantities behaves at $T_{c2}$ as it would at the fixed point of an ordinary
continuous transition, where the local exponents would be size independent
and $U_4$ constant. What the finite lattices display at $T_{c2}$ is a
susceptibility that diverges with $N$ together with a magnetization that
vanishes with $N$, but so slowly that at every accessible size the system
looks ordered.

Where the finite lattices \emph{do} display fixed-point behavior is shown in
Figs.~\ref{fig:peak_height_fss}(c) and (d). The local exponent
$\beta_{\mathrm{eff}}(T)$ of the total magnetization, plotted against
temperature for the five size intervals, fans out on both sides of
$T\simeq1.68$: below it the exponent decreases with size, as in an ordered
phase; above it the exponent increases with size toward the paramagnetic
value $1/2$, which is reached by $T=2.4$; and in $1.66\le T\le1.70$ the five
intervals agree, at $\beta_{\mathrm{eff}}=0.22(2)$. The susceptibility
exponent over the same interval is $\gamma_{\mathrm{eff}}=0.57(4)$, size
independent to within its scatter. Both effective exponents are compatible with the mean-field finite-size values $1/4$ and $1/2$, within the observed finite-size uncertainties. They also
satisfy the hyperscaling relation that $N$-based finite-size scaling imposes,
$2\beta_{\mathrm{eff}}+\gamma_{\mathrm{eff}}=1$, which follows from
$2\beta+\gamma=\nu d$: the measured combination is $1.01(6)$. Read through
$\beta_{\mathrm{eff}}=\beta/\nu_{\mathrm{eff}}$ and
$\gamma_{\mathrm{eff}}=\gamma/\nu_{\mathrm{eff}}$ with the mean-field $\beta=1/2$
and $\gamma=1$, the two exponents give $\nu_{\mathrm{eff}}=2.27(21)$ and
$1.75(12)$, or $1.89(11)$ combined---consistent with the mean-field
$\nu_{\mathrm{eff}}=\nu d=2$. Section~\ref{sec:scaling_variable} shows that the
\emph{width in temperature} of the same transition region does not scale with
$N$ in the same way, and that the discrepancy identifies the correct scaling
variable. The Binder cumulants of the six sizes cross in the same
interval, $1.64\le T\le1.70$, which is where the extrapolated crossings of
Sec.~\ref{sec:binder_analysis} accumulate for bulk, total and core:
$1.638(32)$, $1.663(20)$ and $1.634(58)$. The order-parameter distribution of
these finite systems therefore has an effective fixed point, and it does not lie at
$T_{c2}$. Combining the interval in which the local exponents are size
independent with the Binder extrapolations of Sec.~\ref{sec:binder_analysis}
we identify a finite-size pseudocritical scale at
\begin{equation}
T^{\ast}\simeq1.65-1.70\,J/k_{\mathrm{B}} .
\label{eq:Tstar}
\end{equation}

The two scales are what the phase diagram of Ref.~\cite{Wang2025} implies
for a finite lattice whose free boundary fluctuates. Below $T_{c2}$ the
lattice is a ferromagnet in the ordinary sense: the interior orders
spontaneously, and a boundary transition has been located at the same
coupling~\cite{Okunishi2024,Wang2025}. Between $T_{c2}$ and
$T_{\mathrm{pt}}$ lies the intermediate phase, the hyperbolic counterpart of
the Eggarter phase of the Cayley tree~\cite{Eggarter1974}: the interior is
below its own ordering temperature and orders in response to any
symmetry-breaking perturbation of the boundary, although the symmetric
infinite system carries no order, because the symmetric fixed point of the
boundary-to-bulk recursion is unstable there~\cite{Wang2025}. In a Monte
Carlo simulation with free boundaries the perturbation is supplied by the
boundary itself. Its instantaneous net magnetization is amplified by the
interior, and the ordered interior polarizes the boundary in turn, so that a
finite lattice in this phase carries induced order: a bimodal distribution
of $m$ whose lobes narrow only slowly with $N$, which is what
Table~\ref{tab:exponents} and Fig.~\ref{fig:peak_height_fss} show at
$T_{c2}$ and above it. The finite lattices are therefore ordered-looking on
both sides of $T_{c2}$---spontaneously below it, by induction above
it---which is why the order-parameter distribution has no fixed point there,
while the susceptibility, dominated by the boundary spins, peaks at every
size at the temperature at which the boundary itself orders. The scale
$T^{\ast}$ is the temperature at which induced order gives way to the
paramagnet in a lattice of finite depth. Its exponents are the mean-field
values that Ref.~\cite{Wang2025} reports for the paramagnetic-to-intermediate
transition, and its upward drift with $N$ is what the mechanism predicts,
since a deeper interior amplifies the boundary more strongly. The crossings of Sec.~\ref{sec:binder_analysis} decelerate, moving by $0.03$
over a factor of $739$ in $N$, with increments that fall from $0.014$ to
$0.003$ in a stable geometric ratio of $0.67$; the remaining tail sums to
$0.008$, placing $T^{\ast}(\infty)$ near $1.66$ and far below
$T_{\mathrm{pt}}=2.799$. For the sequence to reach $T_{\mathrm{pt}}$ the
increments would have to stop decelerating and grow, contradicting four
consecutive measured intervals. Section~\ref{sec:scaling_variable} reaches the
same conclusion by a different route and shows, in addition, why a data
collapse cannot by itself decide this question.

Two tree estimates bracket $T_{c2}$ and locate $T_{\mathrm{pt}}$. The Bethe
temperature of the tree of coordination four, $2/\ln2=2.885$, and its dual,
$2/\ln4=1.443$, reproduce $T_{\mathrm{pt}}$ and $T_{c2}$ to $3\%$ and
$2.5\%$~\cite{Wang2025}; and the temperature at which the susceptibility of
the whole tree, boundary included, diverges without spontaneous
magnetization~\cite{Matsuda1974,MullerHartmannZittartz1974,MoritaHoriguchi1975},
$2/\ln(2+\sqrt3)=1.519$, lies $3\%$ above $T_{c2}$, with $2+\sqrt3$ the
growth ratio of the lattice (Sec.~\ref{sec:model}). We do not push these
coincidences further: on the tree the intermediate-to-ferromagnetic
transition is absent, since $K_{c2}\to\infty$ as $p\to\infty$~\cite{Wang2025},
and $\chi$ is monotonic in $T$.

\subsection{The Binder crossing locates the pseudocritical scale $T^{\ast}$, not $T_{c2}$}
\label{sec:binder_analysis}

The fourth-order Binder cumulant,
\begin{equation}
U_4^R(T,L) = 1 - \frac{\langle m_R^4 \rangle}{3\,\langle m_R^2 \rangle^2},
\label{eq:Binder_def}
\end{equation}
is dimensionless, and in a Euclidean system its curves for different sizes
intersect at $T_c$ in a size-independent point. Here the curves intersect, in
all four regions, but not at $T_{c2}$: they intersect at the second scale
identified in Sec.~\ref{sec:peak_height}, and the purpose of this subsection
is to make that statement quantitative.

Two technical points must be settled before the crossings can be read. First,
on a grid of $\Delta T=0.01$ the difference $U_4(L)-U_4(L')$ changes sign
several times in the immediate neighborhood of a genuine intersection, and at
high temperature, where both cumulants fluctuate about zero, it changes sign
tens of times. We therefore group contiguous sign changes, fit a straight line
to the difference over $\pm0.08$ in $T$ about each group, and retain the group
whose slope is most significant relative to its own uncertainty: a physical
intersection has a well-determined nonzero slope, spurious ones do not. The
crossings reported below are unchanged, to within $0.0008$, when the search
window is widened from $[1.30,1.95]$ to $[1.30,2.40]$.

Second, crossings between consecutive sizes become progressively worse
conditioned as $L$ grows, because $U_4(L)$ and $U_4(L+1)$ approach each other
and the intersection angle closes. The bootstrap uncertainty of the bulk
crossing grows from $\pm0.0047$ for the pair $(4,5)$ to $\pm0.0613$ for $(7,8)$
and $\pm0.0782$ for $(8,9)$---an order of magnitude, and comparable to the
entire drift being measured. Crossings against a fixed reference size, which we
take to be $L=4$, keep the curves well separated and yield smooth, decelerating
sequences; these are what we report.

\begin{table}[t]
\centering
\caption{Binder crossing temperatures $T_\times$ and crossing values
$U_4^\ast$ obtained against the fixed reference $L=4$, for the four regions.
Parenthesized digits are bootstrap uncertainties in the last quoted digits.
The sequences move upward, away from $T_{c2}=1.479$, and decelerate; the
extrapolated limits of bulk, total and core agree with one another and
define $T^{\ast}$, inside the interval $[1.479,2.799]$ of the intermediate
phase.}
\label{tab:binder}
\begin{tabular}{lcccc}
\toprule
$(4,L)$ & bulk & boundary & total & core\\
\midrule
$(4,5)$ & 1.5074(47) & 1.6924(89) & 1.6145(74) & 1.3999(37)\\
$(4,6)$ & 1.5331(26) & 1.7082(70) & 1.6282(45) & 1.4328(23)\\
$(4,7)$ & 1.5553(22) & 1.7101(83) & 1.6369(39) & 1.4643(18)\\
$(4,8)$ & 1.5674(22) & 1.7263(87) & 1.6448(41) & 1.4856(12)\\
$(4,9)$ & 1.5799(21) & 1.7338(132)& 1.6476(45) & 1.5015(13)\\
\midrule
$T_\times(\infty)$ & 1.638(32) & 1.761(23) & 1.663(20) & 1.634(58)\\
$U_4^\ast(\infty)$ & 0.506(16) & 0.297(23) & 0.419(14) & 0.573(11)\\
\bottomrule
\end{tabular}
\end{table}

Table~\ref{tab:binder} collects the result. In every region the crossing
temperature increases monotonically with $L$ and moves \emph{away} from
$T_{c2}$; over a factor of $739$ in $N$ the total-system crossing moves from
$1.6145(74)$ to $1.6476(45)$, with increments that fall from $0.014$ to
$0.003$. The crossings therefore do not measure $T_{c2}$: no sequence
approaches it, and the susceptibility peak, which sits on $T_{c2}$ at every
size, lies $0.13$--$0.17$ below every crossing. Extrapolation gives
$1.638(32)$, $1.663(20)$ and $1.634(58)$ for bulk, total and core, three
values that agree within their uncertainties. The boundary sequence
extrapolates higher, to $1.761(23)$, but its cumulant is the flattest of the
four and its crossings the least well conditioned, and its consecutive-pair
crossings, $1.90(8)$, $1.76(8)$ and $1.67(9)$ for $(6,7)$, $(7,8)$ and
$(8,9)$, descend toward the common value. We note that the extrapolation uncertainty is a spread over heuristic parameter choices, so we report $T^*$ as a consistency interval $\simeq 1.65-1.70$ and summarize it as $T^{\ast}=1.67(3)$.

What the crossing measures follows from Sec.~\ref{sec:peak_height}. At
$T^{\ast}$ the local exponents of $\langle|m|\rangle$ and $\chi$ are size
independent and take the mean-field values, and on either side of it the
finite lattices scale as an ordered and as a paramagnetic phase
respectively: $T^{\ast}$ is the fixed point of the order-parameter
distribution of the finite lattices, which is precisely what a Binder
crossing detects. It lies inside $[T_{c2},T_{\mathrm{pt}}]=[1.479,2.799]$,
the intermediate, boundary-sensitive phase identified for this lattice under
open boundary conditions~\cite{Wang2025}, in which bulk order is induced by
the boundary rather than arising from spontaneous symmetry breaking. The slow
upward drift of the crossings is then the finite-size expression of the
instability that defines the intermediate phase: a deeper interior amplifies
the boundary's fluctuations more strongly, so the temperature at which
induced order gives way to the paramagnet moves upward with $N$. 

The crossing values $U_4^\ast$ depend on the region, from $0.297(23)$ at the
boundary to $0.573(11)$ in the core. The boundary value lies $1.2$ standard
deviations from the mean-field $U^\ast=0.27052$, which is consistent with the
mean-field exponents found at $T^{\ast}$, while the core value is close to
the two-dimensional Ising value $0.61069$. We read neither as evidence of a
universality class. The cumulant of a region is the cumulant of a subsystem
magnetization, and it is known to depend on the geometry of the subsystem
and on the boundary condition even in Euclidean
systems~\cite{Selke2006}; on a lattice whose boundary is extensive the four
regions are four different geometries, and a region-dependent $U_4^\ast$ is
what should be expected. The quantity common to the regions is $T^{\ast}$;
the amplitude is not.

\subsection{The scaling variable is the generation index, and the exponent
does not converge}
\label{sec:scaling_variable}

Sections~\ref{sec:peak_height} and~\ref{sec:binder_analysis} take $N$ as the
finite-size scaling variable, following Ref.~\cite{BotetJullienPfeuty1982}, and
locate $T^{\ast}$ from crossings rather than from a data collapse. Both choices
can be tested against the same data. The test returns one positive result, that
the scaling variable is the generation index and not the number of sites, and
one negative result, that no exponent can be extracted from it; the second is
what settles the status of $T^{\ast}$.

Write the scaling ansatz for the Binder cumulant as
\begin{equation}
U_4(T,N) = \mathcal{F}\!\left[(T-T^{\ast})\,N^{1/\nu_{\mathrm{eff}}}\right].
\label{eq:collapse_N}
\end{equation}
The exponent beside $N$ is not the correlation-length exponent. In a Euclidean
system $N=L^{d}$, so $L^{1/\nu}=N^{1/(d\nu)}$ and $\nu_{\mathrm{eff}}=d\nu$;
the mean-field reference for Eq.~\eqref{eq:collapse_N} is therefore
$\nu_{\mathrm{eff}}=2$, not $\nu=1/2$. It is the same $\nu d=2$ that produces
the $\chi^{\max}\propto N^{1/2}$ and $\langle|m|\rangle\propto N^{-1/4}$ used
as reference in Sec.~\ref{sec:peak_height}, and that the exponents measured at
$T^{\ast}$ reproduce.

\begin{figure*}[t]
\centering
\includegraphics[width=\textwidth]{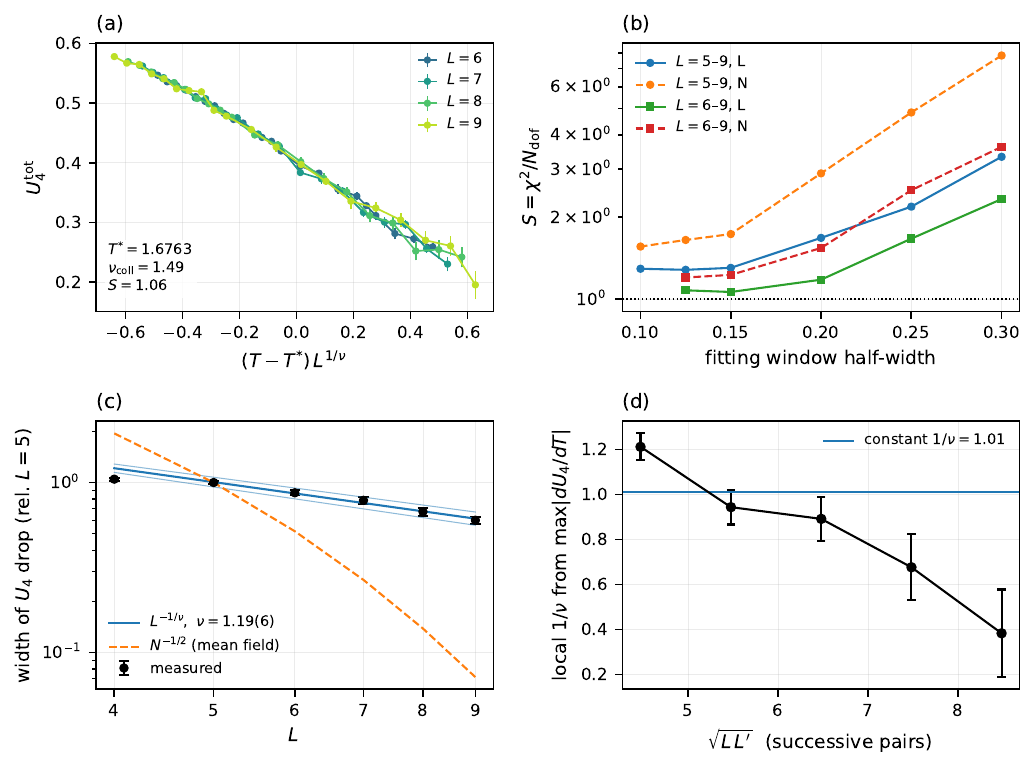}
\caption{Finite-size scaling of the Binder cumulant in the generation index.
(a) Superposition of $U_4^{\mathrm{tot}}$ for $L=6$--$9$ in the variable
$(T-T^{\ast})L^{1/\nu}$, over a fitting window of half-width $0.15$ about
$T^{\ast}$; the fitted values and the collapse statistic are quoted in the
panel. (b) $S=\chi^2/N_{\mathrm{dof}}$ against the half-width of the fitting
window, for two sets of sizes and for both candidate scaling variables. $S$
falls as the window narrows and then stabilizes at $S=1.10(9)$ over the five
narrowest windows of the $L=6$--$9$ data in the variable $L$. Scaling in $L$
lies below scaling in $N$ at every window. (c) Width in temperature over which
$U_4^{\mathrm{tot}}$ falls from $0.55$ to $0.15$, relative to $L=5$, on
logarithmic axes; the solid line is the power law fitted to $L\ge5$ with its
one-standard-deviation band and the dashed line the mean-field prediction
$N^{-1/2}$, which fails by a factor $8.4$ at the largest size. The $L=4$ point
lies above the fitted law and is excluded from the analysis; where not
visible, error bars are smaller than the symbols. (d) Local exponent
obtained from the height of the peak of $|dU_4/dT|$ between successive sizes,
an estimator in which $T^{\ast}$ does not appear. It is not constant: it falls
monotonically across the whole range, and a constant is excluded with
$\chi^2=29.1$ for four degrees of freedom. Panels (a)--(c) show that the curves
of different size can be superposed; panel (d) shows that the factor by which
they must be stretched does not converge to a power of $L$. The two statements
are not in conflict, and the combination is the result: superposition whose
required stretch is still drifting at the largest size is the signature of a
crossover rather than of a fixed point. The collapse in (a) is not thereby
spurious---(b) shows $S$ stabilizing rather than continuing to fall as the
window narrows---but $T^{\ast}$ and $\nu$ are correlated there at
$\rho=+0.84$, which is why the exponent is read from the $T^{\ast}$-free
estimator of (d) and not from (a).}
\label{fig:scaling_variable}
\end{figure*}

That reference is excluded by the temperature width of the transition region.
Table~\ref{tab:widths} and Fig.~\ref{fig:scaling_variable}(c) give the interval
in $T$ over which $U_4^{\mathrm{tot}}$ falls from $0.55$ to $0.15$, an
estimator that requires no fit. Under Eq.~\eqref{eq:collapse_N} this width
scales as $N^{-1/\nu_{\mathrm{eff}}}$, so between $L=5$ and $L=9$, across a
factor $195$ in $N$, mean-field scaling predicts a narrowing by a factor
$0.072$. The measured narrowing is $0.600$: the prediction fails by a factor
$8.4$, far outside anything the error bars or the residual drift of
Table~\ref{tab:exponents} could absorb. Whatever controls the width of the
transition in temperature, it is not $N^{1/2}$.

\begin{table}[t]
\centering
\caption{Temperature width over which $U_4^{\mathrm{tot}}$ falls from $0.55$ to
$0.15$, relative to $L=5$, compared with the mean-field prediction of
Eq.~\eqref{eq:collapse_N}. Scaling in $N$ is excluded by a factor $8.4$ at the
largest size.}
\label{tab:widths}
\begin{ruledtabular}
\begin{tabular}{ccccc}
$L$ & $N$ & width $[J/k_{\mathrm{B}}]$ & measured & $N^{-1/2}$ \\ \hline
4 & 241     & 0.5229(102) & 1.047 & 1.946 \\
5 & 913     & 0.4995(114) & 1.000 & 1.000 \\
6 & 3\,421  & 0.4345(142) & 0.870 & 0.517 \\
7 & 12\,781 & 0.3923(194) & 0.785 & 0.267 \\
8 & 47\,713 & 0.3366(180) & 0.674 & 0.138 \\
9 & 178\,081& 0.2997(141) & 0.600 & 0.072 \\
\end{tabular}
\end{ruledtabular}
\end{table}

On a hyperbolic lattice the two candidate variables are not related by a power.
Here $N\sim\lambda^{L}$ with $\lambda=2+\sqrt{3}$ (Sec.~\ref{sec:model}), so
$N$ is \emph{exponential} in the generation index, which is the number of steps
separating the boundary from the core and the only geometric length the lattice
possesses. If the scaling variable is $L$, a fit performed in $N$ over a finite
range of sizes returns
\begin{equation}
\nu_{\mathrm{eff}} \;=\; \nu\,
\frac{\ln (N_{\max}/N_{\min})}{\ln (L_{\max}/L_{\min})} ,
\label{eq:nu_conversion}
\end{equation}
which for our sizes is a factor between $8.4$ and $9.7$. A fit in $N$ returning
$\nu_{\mathrm{eff}}\simeq9$, a value with no interpretation as a
correlation-length exponent, is then a fit in the wrong variable.

Fitting the widths of Table~\ref{tab:widths} to $L^{-1/\nu}$ identifies the
smallest lattice as an obstruction. Over $L=4$--$9$ the fit is rejected,
$\chi^2/N_{\mathrm{dof}}=3.76$ with $p=0.005$ and residuals showing systematic
curvature rather than scatter; the width of $L=4$ differs from that of $L=5$ by only
$4\%$, while the step from $L=8$ to $L=9$ is $11\%$. With four generations
there is no asymptotic regime to speak of, and we exclude $L=4$ from the
finite-size analyses of this subsection. Excluding it the fit is acceptable,
$\nu=1.19(6)$ with $\chi^2/N_{\mathrm{dof}}=0.29$, and excluding $L=5$ as well
gives $\nu=1.10(8)$ with $\chi^2/N_{\mathrm{dof}}=0.24$. The uncertainty on
each width is obtained by propagating $\sigma(U_4)$ evaluated at each of the
two crossings rather than at their average: $\sigma(U_4)$ grows by a factor
three across the transition region (Appendix~\ref{app:dynamics}), so the
crossing at $U_4=0.15$, on the high-temperature side, carries most of it.

Within that variable the curves can be superposed. Using the statistic
$S=\chi^2/N_{\mathrm{dof}}$ of Houdayer and
Hartmann~\cite{HoudayerHartmann2004}, in which every point is weighted by its
jackknife error and compared with the interpolation of the remaining curves,
$S$ falls as the fitting window is narrowed about $T^{\ast}$ and then
stabilizes: the five windows from a half-width of $0.200$ down to $0.075$ give
$S=1.10\pm0.09$ for $L=6$--$9$ in the variable $L$
[Fig.~\ref{fig:scaling_variable}(b)]. $S$ does not continue to fall, which is
what a spurious improvement from imposing fewer constraints would produce.
Within that window the fit returns $T^{\ast}=1.6607(24)$ for $L=5$--$9$ and
$1.6741(39)$ for $L=6$--$9$ [Fig.~\ref{fig:scaling_variable}(a)], in agreement
with Eq.~\eqref{eq:Tstar} and with the crossings of
Sec.~\ref{sec:binder_analysis}. The variable is identified without
extrapolation: on changing the set of sizes from $L=5$--$9$ to $L=6$--$9$ the
fitted exponent moves by $12$--$15\%$ in the variable $L$ and by $19$--$28\%$
in the variable $N$, in every window.

\begin{table}[t]
\centering
\caption{Height of the maximum of $|dU_4/dT|$ for the total system, obtained by
differentiating a four-parameter logistic fitted to $U_4^{\mathrm{tot}}$ over
$|T-1.67|\le0.25$, with bootstrap uncertainties; a local cubic fit gives the
same values to within one standard deviation. The last column is the local
exponent between successive sizes. A constant, which is what a fixed point
would require, is excluded with $\chi^2=29.1$ for four degrees of freedom,
$p=8\times10^{-6}$; the fall from first to last interval is $4.1$ standard
deviations.}
\label{tab:peak_slope}
\begin{ruledtabular}
\begin{tabular}{ccc}
$L$ & $\max|dU_4/dT|$ & local $1/\nu$ \\ \hline
4 & 0.706(7)  &           \\
5 & 0.925(9)  & 1.211(62) \\
6 & 1.099(12) & 0.945(80) \\
7 & 1.261(14) & 0.892(101)\\
8 & 1.380(21) & 0.675(141)\\
9 & 1.443(24) & 0.379(191)\\
\end{tabular}
\end{ruledtabular}
\end{table}

What the superposition does not establish is an exponent, and this is the
central point of the subsection. Two facts stand in the way. The first is that
in the collapse $T^{\ast}$ and $\nu$ are correlated at $\rho=+0.84$, so
narrowing the window trades one against the other and the fitted exponent
drifts by $19\%$ across the windows of Fig.~\ref{fig:scaling_variable}(b); the
widths, which carry no such degeneracy, give $\nu=1.19(6)$, some $20\%$ lower
and two standard deviations away. The
second is decisive. The height of the maximum of $|dU_4/dT|$ scales as
$L^{1/\nu}$ at a fixed point, and $T^{\ast}$ does not enter it, so the
degeneracy is absent by construction. Table~\ref{tab:peak_slope} and
Fig.~\ref{fig:scaling_variable}(d) show that this estimator does not return a
constant exponent: the local value falls monotonically from $1.211(62)$ to
$0.379(191)$, a constant is excluded with $\chi^2=29.1$ for four degrees of
freedom, and a single power law fitted to the heights gives
$\chi^2/N_{\mathrm{dof}}=5.1$ with $p=0.0016$. A saturating form
$A-BL^{-z}$ and a logarithm describe the same heights with
$\chi^2/N_{\mathrm{dof}}=1.01$ and $1.21$; with six sizes we cannot distinguish
between them, and the distinction does not matter here, because neither is a
power-law divergence.

Two instrumental explanations for this have been excluded by direct
measurement, and both are reported in Appendix~\ref{app:dynamics}. Incomplete
equilibration would flatten the cumulant at the largest size, since
$N_{\mathrm{eq}}$ is fixed while $\tau_{\mathrm{int}}$ grows with $L$; running
$L=9$ from ordered and from random initial configurations and comparing
$U_4^{\mathrm{tot}}$ as a function of the number of steps discarded shows
agreement to within $0.4$--$2.5$ standard deviations and no residual drift.
Misestimated error bars would distort the weighted fits; cutting long chains
into segments of $4\times10^{4}$ steps and analyzing each as a production run
gives quoted errors and observed scatter consistent at $p=0.41$.

The two halves of this subsection are therefore consistent, and together they
say something sharper than either alone. Over the range of sizes we can reach,
the cumulant curves are superposable in the generation index, with a quality
that a genuine scaling function would give. But the factor by which they must
be stretched does not approach a power of $L$: it drifts, and the drift is
still under way at the largest lattice. A family of smooth curves related by a
slowly varying stretch is what a crossover looks like, not what a critical
point looks like. We therefore report the scaling variable, which the data fix,
and not an exponent, which they do not; and this is the same conclusion that
Sec.~\ref{sec:peak_height} reaches from the drift of $\beta_{\mathrm{eff}}$ and
of the Binder crossings, reached here through an independent observable.

One further property of Eq.~\eqref{eq:collapse_N} is worth recording, because
it bears on how $T^{\ast}$ should be located at all. Consider a size-dependent
shift $T^{\ast}(N)\to T^{\ast}(N)+c\,N^{-1/\nu_{\mathrm{eff}}}$. Then
\begin{equation}
\bigl[T-T^{\ast}(N)-c\,N^{-1/\nu_{\mathrm{eff}}}\bigr]N^{1/\nu_{\mathrm{eff}}}
=\bigl[T-T^{\ast}(N)\bigr]N^{1/\nu_{\mathrm{eff}}}-c ,
\label{eq:flat_direction}
\end{equation}
so the abscissa of \emph{every} curve is displaced by the same constant and the
scaling function absorbs it, $\mathcal{F}(x)\to\mathcal{F}(x+c)$. The
invariance is exact, and we verify numerically that $S$ is unchanged to all
printed digits along this one-parameter family. A collapse therefore determines
$T^{\ast}(N)$ only up to Eq.~\eqref{eq:flat_direction}. In a Euclidean analysis
the degeneracy is closed by imposing a single size-independent $T^{\ast}$,
which is precisely the hypothesis under test here; and the gauge freedom decays
only as $N^{-1/\nu_{\mathrm{eff}}}$, a factor that falls by half across the two
and a half decades in $N$ available to us, so that within our range even the
sign of the drift can be reversed at unchanged collapse quality. This is a
property of the ansatz rather than of our statistics, and it is why we locate
$T^{\ast}$ from the Binder crossings: the crossing condition fixes the gauge by
a physical requirement---curves meeting at a common value of the
cumulant---instead of leaving it free.

\subsection{The upper transition is not accessible with open boundaries}
\label{sec:upper_transition}

The published phase diagram places a second transition at
$T_{\mathrm{pt}}=2.799$~\cite{Wang2025,Krcmar2008}, between the paramagnetic
and the intermediate phase. Our temperature grid covers it, with a refinement
of $\Delta T=0.05$ over $[2.65,3.00]$. We find no trace of it.

For $T\ge2.4$ and for $L=7,8,9$, $\chi_R$ decreases strictly monotonically
point by point in all four regions: there is no maximum, no shoulder and no
inflection anywhere near $2.799$. Over the same range $|U_4^R|\le0.045$ in the
worst case and $\le0.03$ typically, that is, the cumulant is zero within its
scatter, the value for a Gaussian order parameter. And the susceptibility
itself is size independent there: at $T=2.80$ the bulk values are $0.638$,
$0.632$ and $0.647$ for $L=7,8,9$, unchanged to within $3\%$ while $N$
increases by a factor of $14$. A transition would show growth with $N$; there
is none.

The absence is not a failure of resolution, and the deep core does not help. It
follows from what the transition at $T_{\mathrm{pt}}$ is. Under a free
boundary, $T_{\mathrm{pt}}$ is the temperature below which the interior of
the infinite lattice becomes susceptible to a symmetry-breaking perturbation
of the boundary~\cite{Wang2025}. The order that this susceptibility induces
in a lattice of $L\le9$ generations is the induced order of
Sec.~\ref{sec:peak_height}, and it is visible only up to
$T^{\ast}(N)\simeq1.65$, because a lattice of finite depth amplifies its
boundary too weakly above that temperature. The
qualification matters, and Sec.~\ref{sec:perturbed_boundary} makes it precise:
what fails above $T^{\ast}$ is the amplification of the \emph{incoherent}
fluctuation of a free boundary, whose coherent component is
$O(N_{\partial}^{-1/2})$, not the amplification of a coherent perturbation,
which survives up to $T_{\mathrm{pt}}$. Between $T^{\ast}$ and
$T_{\mathrm{pt}}$ the finite lattices are paramagnetic, and the region
averages, dominated by boundary spins, carry no signature of
$T_{\mathrm{pt}}$. The CTMRG determinations of $T_{\mathrm{pt}}$ are obtained
under a fixed or perturbed boundary and by measuring at the center of the
system~\cite{Krcmar2008,Wang2025}, a point emphasized by Asaduzzaman
\textit{et al.} in the context of open boundaries and
holography~\cite{Asaduzzaman2022}. That is a different observable, and our
result does not contradict it.

\subsection{The upper transition under a perturbed boundary}
\label{sec:perturbed_boundary}

Section~\ref{sec:upper_transition} leaves open whether $T_{\mathrm{pt}}$ is
inaccessible to a Monte Carlo simulation as such or only to the observable
measured there. The phases of Ref.~\cite{Wang2025} are defined by the response
of deep-in-bulk observables to an infinitesimal symmetry-breaking perturbation
of the boundary: the induced magnetization vanishes only in the paramagnetic
phase, and the intermediate phase is distinguished from the ferromagnetic one by
the sensitivity of the nearest-neighbor correlation function to that
perturbation. Both criteria can be implemented on a finite lattice by adding to
Eq.~\eqref{eq:Ising_H} a field acting on the outermost generation only,
$-h\sum_{i\in\partial}\sigma_i$, and resolving the response generation by
generation.

Two features of the finite lattice fix how this has to be done. First, region
averages are again the wrong observable. The outermost generation of the core
holds $73\%$ of its sites and lies two generations from the boundary at every
$L$, so the core magnetization at fixed $h$ is set by the response at depth two
and is nearly independent of $L$ on both sides of $T_{\mathrm{pt}}$. Second,
below $T^{\ast}(N)$ the zero-field distribution of $m$ is bimodal, and a field
selects a sector as soon as $\beta hN_{\partial}\langle|m_{\partial}|\rangle
\gtrsim1$, a condition that is met more easily the larger the lattice; growth of
the response with $L$ at fixed $h$ is therefore not, by itself, evidence of
boundary-induced order.

We therefore measure the magnetization $m(d)$ of the generation at depth $d$
from the boundary, with $d=0$ the boundary itself, and the amplification factor
\begin{equation}
\Lambda(d) = \frac{\langle m(d)\rangle}{\langle m(d-1)\rangle}.
\label{eq:Lambda}
\end{equation}
In linear response on the Cayley tree, away from the boundary layer
and from the center, $\Lambda(d)$ approaches the eigenvalue $(q-1)\tanh K$ of
the linearized boundary-to-bulk recursion of Ref.~\cite{Wang2025}: it exceeds
unity where the zero fixed point is unstable and falls below unity where the
interior attenuates the boundary signal. The crossing $\Lambda(d)=1$ therefore
locates $T_{\mathrm{pt}}$ without a finite-size scaling ansatz, provided three
conditions hold, all of which are tested: $\Lambda(d)$ is independent of $L$ at
fixed $d$ and $h$; its crossing depends on the field only through the $O(h^2)$
term that $Z_2$ parity permits, so that the $h\to0$ limit exists and is reached
along a known form; and that limit does not drift with $d$ once the boundary
layer is left behind. The second condition is the one that has to be measured
rather than assumed, and it is not the statement that the crossing is
independent of $h$: at the strongest field used here it is displaced by several
times its own statistical error.

The simulations use purely local Metropolis updates. We use $L=6$, $7$ and $8$, eleven temperatures between
$2.40$ and $3.20$ with a spacing of $0.05$ over $[2.60,3.00]$, and three field
strengths, $h=0.03$, $0.05$ and $0.1$ in units of $J$; each point is
$5\times10^{3}$ MCS of thermalization and $2\times10^{5}$ MCS of measurement,
with the generation-resolved profile recorded every ten sweeps. 

Before applying the estimator to the hyperbolic lattice we calibrated it on a
case where the answer is known in closed form. On the Cayley tree of the same
coordination $q=4$, the recursion above is exact, $\Lambda=(q-1)\tanh K$, and
the threshold is $T=1/\mathrm{arctanh}(1/3)=2.88539$. Simulating that tree with
nine generations and the same kernel, observable, fitting procedure and field
$h=0.1$, the measured $\Lambda(d=2)$ reproduces $3\tanh(1/T)$ point by point
within $1.2$ standard deviations, and the crossing gives $2.8864(66)$ at $d=2$
and $2.8848(114)$ at $d=3$, unbiased at the level of $0.05\%$. The bias appears
only when the generation used becomes small: $-0.68\%$ at $d=4$, where that
generation holds $108$ sites, and $-3.1\%$ at $d=5$, where it holds $36$. The
estimator is therefore reliable while the generations entering the ratio hold
some hundreds of sites, and it is biased downward on approach to the center.
Accordingly we quote $d=2$ and $d=3$ and use $d=4$, which at $L=8$ is the
generation of $180$ sites, as a consistency check only.

\begin{table}[t]
\centering
\caption{Temperature at which the amplification factor of
Eq.~\eqref{eq:Lambda} crosses unity, for three depths $d$ from the boundary,
three sizes and three field strengths. Uncertainties are from the weighted
straight-line fit through the crossing, inflated by $\sqrt{\chi^2/\nu}$ where
that exceeds unity. The $d=4$ column is a consistency check: the Cayley-tree
calibration shows the estimator biased low by $0.7\%$ at the corresponding
generation size.}
\label{tab:Tpt}
\begin{tabular}{ccccc}
\toprule
$h$ & $L$ & $d=2$ & $d=3$ & $d=4$\\
\midrule
$0.03$ & 6 & 2.827(13) & 2.839(34) & 2.717(51)\\
       & 7 & 2.823(9)  & 2.796(12) & 2.787(33)\\
       & 8 & 2.809(4)  & 2.817(8)  & 2.799(14)\\
$0.05$ & 6 & 2.821(9)  & 2.812(11) & 2.869(51)\\
       & 7 & 2.808(5)  & 2.797(21) & 2.803(19)\\
       & 8 & 2.811(4)  & 2.812(8)  & 2.819(18)\\
$0.1$  & 6 & 2.802(4)  & 2.797(11) & 2.801(28)\\
       & 7 & 2.811(2)  & 2.800(5)  & 2.796(8)\\
       & 8 & 2.808(2)  & 2.799(2)  & 2.793(4)\\
\bottomrule
\end{tabular}
\end{table}

\begin{figure}[t]
\centering
\includegraphics[width=\linewidth]{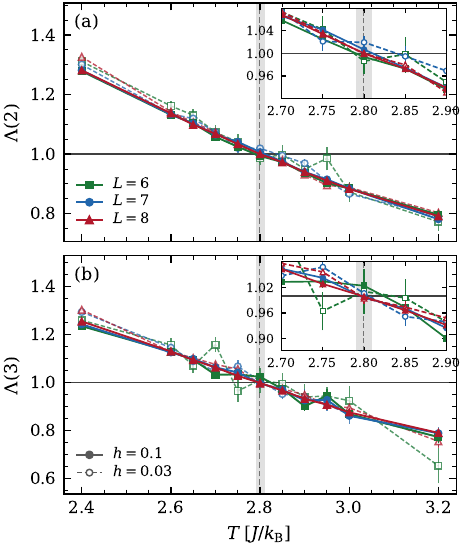}
\caption{Amplification factor of Eq.~\eqref{eq:Lambda} against temperature, at
depth (a) $d=2$ and (b) $d=3$ from the boundary, for the three sizes and the
two extreme field strengths; filled symbols and solid lines are $h=0.1$, open
symbols and dashed lines $h=0.03$. The intermediate field $h=0.05$ is omitted
here for legibility, its curves lying between the two shown; it appears in
Table~\ref{tab:Tpt} and in Fig.~\ref{fig:extrap_h}. Above $\Lambda=1$ each generation carries more
induced magnetization than the one outside it and the lattice amplifies the
boundary inward; below it the interior attenuates the boundary signal. The
curves for the three sizes coincide within their uncertainties at fixed depth
and field, which is the condition that makes the crossing a property of the
lattice rather than of its size. The vertical band is
$T_{\mathrm{pt}}=2.81(1)$ from Eq.~\eqref{eq:Tpt} and the dashed vertical line
the corner-transfer-matrix value $2.799$ of Ref.~\cite{Wang2025}. Insets:
the crossing region enlarged.}
\label{fig:Lambda}
\end{figure}

Figure~\ref{fig:Lambda} shows the measured profiles and
Table~\ref{tab:Tpt} collects the twenty-seven crossings. All lie between $2.72$
and $2.87$, and the three conditions stated above are met. Independence of $L$
at fixed depth and field holds with $\chi^2/\nu$ between $0.03$ and $1.97$; the
weighted means at $d=2$ are $2.812(3)$ at $h=0.03$, $2.811(3)$ at $h=0.05$ and
$2.809(1)$ at $h=0.1$.

\begin{figure}[t]
\centering
\includegraphics[width=\linewidth]{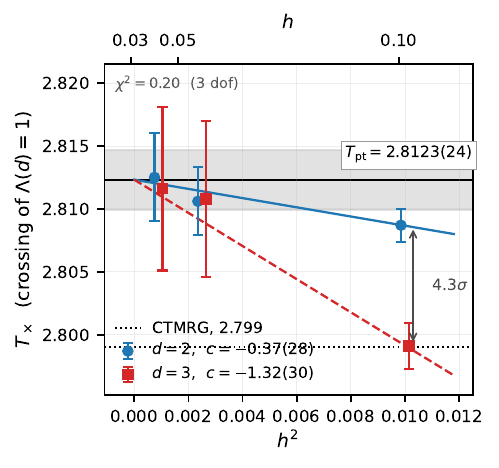}
\caption{Extrapolation of the crossing $\Lambda(d)=1$ to zero field. Each point
is the weighted mean over $L=6$, $7$ and $8$ at fixed depth and field, from
Table~\ref{tab:Tpt}; the abscissa is $h^2$, the variable in which $Z_2$ parity
requires the crossing to be linear, and the upper axis gives the corresponding
$h$. Lines are the joint fit of Eq.~\eqref{eq:h_joint}, in which the curvature
is free for each depth but the intercept is shared. The band is the fitted
$T_{\mathrm{pt}}$ with its statistical error and the dotted line the
corner-transfer-matrix value of Ref.~\cite{Wang2025}. The $4.3\sigma$ separation
between the two depths at the strongest field is the whole of the drift with
depth, and it closes as $h\to0$. Points are offset horizontally for legibility;
the fit uses the true abscissae.}
\label{fig:extrap_h}
\end{figure}

The form of the $h\to0$ extrapolation is fixed by symmetry rather than chosen.
Under $h\to-h$ every magnetization changes sign, so the ratio
$\Lambda(d)$ of Eq.~\eqref{eq:Lambda} is invariant: $\Lambda$ is an
\emph{even} function of $h$, its expansion is
$\Lambda=\Lambda_0+\Lambda_2h^2+O(h^4)$, and the temperature at which it
crosses unity moves as
\begin{equation}
T_\times(h) = T_{\mathrm{pt}} + c\,h^{2} + O(h^{4}).
\label{eq:h_squared}
\end{equation}
The three fields measured over-determine Eq.~\eqref{eq:h_squared}, and they
test more of it than the quadratic form alone. The coefficient $c$ is a term in
the expansion of $\Lambda(d)$ and must therefore depend on the depth at which
that ratio is taken; $T_{\mathrm{pt}}$ is the transition temperature of the
lattice and must not. Fitting the six weighted means at $d=2$ and $d=3$ with a
single intercept and one curvature per depth,
\begin{equation}
T_\times(d,h) = T_{\mathrm{pt}} + c_d\,h^{2},
\label{eq:h_joint}
\end{equation}
gives $T_{\mathrm{pt}}=2.8123(24)$, $c_2=-0.37(28)$ and $c_3=-1.32(30)$, with
$\chi^{2}=0.20$ for three degrees of freedom (Fig.~\ref{fig:extrap_h}). Each
half of that prediction can then be tested against the alternative it excludes.
Forcing one curvature for both depths raises $\chi^{2}$ to $18.7$ for four
degrees of freedom, a change of $18.5$ for a single parameter: the curvature
does depend on depth, as it must. Removing the field dependence altogether
gives $\chi^{2}=26.0$ for five, so the $h^{2}$ term is required by the data and
not merely permitted by it. Allowing instead a separate intercept for each
depth improves the fit by $\Delta\chi^{2}=0.07$ for one further parameter and
returns $2.8120(26)$ at $d=2$ against $2.8136(54)$ at $d=3$, which agree to
$0.3$ standard deviations: the intercept is common, as it must be.

That last comparison settles the third of the three conditions quantitatively,
and it does so at the one place where the raw data appear to violate it. At
$h=0.1$ the crossings at the two calibrated depths differ by $0.0096(22)$,
which is $4.3$ standard deviations; at $h=0.05$ and $h=0.03$ the same
difference is $-0.0002(68)$ and $0.0009(74)$. The apparent drift with depth is
thus a property of the strongest field rather than a residual bias of the
estimator, it carries the depth dependence that Eq.~\eqref{eq:h_squared}
requires of it, and it is absent in the limit. One caveat on the quality of the
joint fit: $\chi^{2}=0.20$ for three degrees of freedom is low because the
crossing uncertainties of Table~\ref{tab:Tpt} are deliberately conservative,
the covariance of the straight-line fit being already scaled by its own
residual before the further $\sqrt{\chi^2/\nu}$ inflation quoted there. The
error on the extrapolated intercept inherits that conservatism.

We note that decreasing $h$ indefinitely is counterproductive rather than
conservative: the induced profile falls linearly with $h$ while the statistical
noise on $m(d)$ remains $O(N_d^{-1/2})$, so below some field the estimator is
noise dominated. Equation~\eqref{eq:h_squared} makes the extrapolation
available without going there. A stronger version of the first condition
is that the whole curve $\Lambda(T)$, and not merely its root, should collapse
across sizes at fixed $d$ and $h$: within the fitting window it does, with
$\chi^2/\nu\le1.5$ in every case.

Restricting to the calibrated depths leaves eighteen determinations, spanning
$2.796$ to $2.839$ with a mean of $2.8105$ and a standard deviation of
$0.0116$, and the zero-field intercept of Eq.~\eqref{eq:h_joint} is
$2.8123(24)$. We quote
\begin{equation}
T_{\mathrm{pt}} = 2.81(1)\,J/k_{\mathrm{B}},
\label{eq:Tpt}
\end{equation}
where the uncertainty is the spread over depths, sizes and fields and not the
statistical error of the extrapolation, which is four times smaller. The
distinction matters here more than the arithmetic suggests. What the Cayley
tree calibrates is the \emph{estimator}: that the profile $m(d)$, its ratio and
the straight-line fit through the crossing return the threshold of a system
whose threshold is known. It does not calibrate the \emph{criterion}. The
identification of $\Lambda(d)=1$ with $T_{\mathrm{pt}}$ rests on a
generation-to-generation recursion that closes exactly only on a tree, and the
$\{5,4\}$ lattice has loops both within a generation and between consecutive
ones. Nothing in this subsection measures the residual of that approximation.
Our extrapolated intercept lies $0.48\%$ above the corner-transfer-matrix
result $T_{\mathrm{pt}}=1/K_{c1}=2.799$ of Ref.~\cite{Wang2025}, a displacement
comparable to the spread of the determinations themselves and an order of
magnitude larger than the $0.05\%$ at which the estimator is calibrated. We
therefore read the agreement at the level of that spread, which is what
Eq.~\eqref{eq:Tpt} states. Reading it instead at the level of the statistical
error of the extrapolation would turn it into a $5.5\sigma$ discrepancy, and
that would be an artefact of quoting an uncertainty that accounts for the
estimator while the leading systematic lies in the criterion.

Two remarks on what this does and does not establish. It is not a finite-size
scaling determination: no quantity is extrapolated in $N$, and the agreement
across $L=6$, $7$ and $8$ is a consistency check rather than an extrapolation. And it does not
contradict Sec.~\ref{sec:upper_transition}: with a free boundary the
perturbation reaching the interior is the incoherent fluctuation of
$N_{\partial}$ spins, whose coherent part is $O(N_{\partial}^{-1/2})$, and the
condition for that to survive $L$ generations of amplification is
$\Lambda/\sqrt{b}>1$ rather than $\Lambda>1$. On the tree with $b=3$ the former
gives $b\tanh^{2}K=1$, that is $T=2/\ln(2+\sqrt{3})=1.5186$, the
Müller-Hartmann--Zittartz temperature already encountered in
Sec.~\ref{sec:peak_height} as the boundary extrapolation of the susceptibility
peak. The two thresholds are the coherent and incoherent versions of the same
recursion, and the free-boundary observables of Secs.~\ref{sec:peak_height}
and~\ref{sec:upper_transition} see only the second.

\section{Conclusions}
\label{sec:conclusions}

We have studied the Ising model on the $\{5,4\}$ hyperbolic lattice with free
boundaries, a geometry in which three out of four spins lie on the boundary at
every size, and asked what can and cannot be measured there with standard Monte
Carlo tools.

We find that in a geometry where the boundary remains macroscopic, different Monte Carlo observables cease to identify the same scale, requiring each observable to be matched with its underlying physical mechanism. The finite lattices are organized by two separate temperature scales. Requiring that the
susceptibility maximum agree across four nested regions of very different
boundary content gives $T_{c2}=1.4782(15)\,J/k_{\mathrm{B}}$ with no
extrapolation, within $0.05\%$ of the CTMRG value~\cite{Wang2025}, and a peak
height consistent with an effective scaling $N^{0.8}$. That temperature scale is not a fixed point of the
order-parameter distribution: instead, a finite-size pseudocritical scale emerges at $T^{\ast}\simeq 1.67$,
inside the boundary-sensitive intermediate phase, where the Binder curves cross
and effective exponents are compatible with the $1/4$ and $1/2$ mean-field values. Both scales follow from that
phase once the boundary fluctuates, the interval between them being where the
interior amplifies boundary fluctuations into induced order.

The geometry also fixes which variable finite-size scaling must be written in,
and the answer is not the one Euclidean intuition suggests. Because
$N\sim\lambda^{L}$, the number of sites is exponential in the generation index,
and the two are not related by any power. The temperature width of the
transition region settles the question without a fit: mean-field scaling in $N$
predicts that it narrow by a factor $0.072$ between $L=5$ and $L=9$, and the
measured narrowing is $0.600$. Written in the generation index the same data
admit a scaling window, narrow but well defined, in which the collapse
statistic is $S=1.10(9)$ over five nested fitting windows. Read in $N$ this
same behavior returns an exponent near $9$, a number with no interpretation as
a correlation-length exponent and which is simply $\nu\simeq1$ expressed in the
wrong variable. A separate consequence is that a data collapse cannot decide
whether $T^{\ast}$ drifts with size at all: a size-dependent shift
$c\,N^{-1/\nu_{\mathrm{eff}}}$ displaces every curve's abscissa by the same
constant and is absorbed exactly by the scaling function, so the crossing
determination we use is better defined than a collapse rather than a weaker
substitute for one. What the collapse does not yield is an exponent, and the reason is a result in
itself. The height of the maximum of $|dU_4/dT|$ scales as $L^{1/\nu}$ at a
fixed point and does not involve $T^{\ast}$, so it measures the exponent free
of the correlation between the two; measured this way the exponent is not
constant, falling monotonically from $1.211(62)$ to $0.379(191)$ across our
sizes, and a single power law is rejected at $p=0.0016$. Incomplete
equilibration and misestimated error bars are excluded as causes by direct
measurement. Curves of different size that superpose but whose relative stretch
is still drifting at the largest lattice are what a crossover looks like, not a
critical point. We therefore report the scaling variable, which the data fix,
and not an exponent, which they do not---an independent route to the same
conclusion that the drift of $\beta_{\mathrm{eff}}$ and of the Binder crossings
already indicated, that $T^{\ast}$ organizes the finite lattices without being
a fixed point of the thermodynamic limit. 

The upper
thermodynamic transition leaves no trace in any region average of the free-boundary system,
as it should not; it becomes accessible once the boundary is perturbed. The
generation-resolved response to a field on the outermost shell amplifies inward
below $T_{\mathrm{pt}}$ and decays inward above it, and the crossing gives
$T_{\mathrm{pt}}=2.81(1)\,J/k_{\mathrm{B}}$ from eighteen determinations that
agree across three sizes, two depths and three field strengths; the same
estimator is unbiased to $0.05\%$ on the Cayley tree of the same coordination,
where the threshold is known in closed form. The third field turns the
zero-field extrapolation into a test rather than an assumption: $Z_2$ parity
makes the crossing linear in $h^2$ with a depth-dependent slope and a common
intercept, and the six measurements at the two calibrated depths meet that
form with $\chi^2=0.20$ for three degrees of freedom, the only visible drift
with depth being the $O(h^2)$ displacement of the strongest field. Both transitions are therefore
within reach of a simulation of a finite lattice, provided the observable is
matched to the quantity that defines them.

\appendix

\section{Validation of the Monte Carlo scheme}
\label{app:validation}
\label{sec:validation}

Because the lattice construction of Sec.~\ref{sec:model} is nonstandard, we
first verify the simulation machinery on a case with an exact answer, using the
same update kernels, the same jackknife estimators and the same analysis code.
We simulate the Ising model on the square lattice with periodic boundary
conditions at $L=8,16,32$, with $3\times10^{4}$ measurements per temperature on
a grid of spacing $\Delta T=0.025$ over $[2.14,2.44]$.

The Binder crossings give $T_c=2.2612$ for the pair $(8,16)$, $2.2668$ for
$(16,32)$ and $2.2652$ for $(8,32)$, against the exact value
$T_c=2/\ln(1+\sqrt{2})=2.269185$: deviations of $0.35\%$, $0.10\%$ and
$0.18\%$, the deviation falling to $0.10\%$ for the largest pair. The crossing value for
the largest pair is $U_4^\ast=0.6126$, to be compared with the universal
two-dimensional value $0.61069$~\cite{KamieniarzBlote1993}, a deviation of
$0.3\%$. We note that this reference value is specific to periodic boundary
conditions on a square lattice, the geometry of our control run; the critical
cumulant is known to depend sensitively on the boundary condition and on the
region over which the moments are taken~\cite{Selke2006}, a sensitivity that
returns in a more severe form in Sec.~\ref{sec:binder_analysis}.

One further feature of the control run is worth recording because it recurs
with the opposite sign below. Under periodic boundary conditions the
susceptibility maxima approach $T_c$ \emph{from above}, at $T=2.440$, $2.390$
and $2.340$ for $L=8,16,32$. In the hyperbolic lattice with open boundaries the
corresponding shifts run in the opposite direction, and, as we show in
Sec.~\ref{sec:peaks_regions}, in a region-dependent way.

\section{Autocorrelation, equilibration and error bars}
\label{app:dynamics}

The static validation of Appendix~\ref{app:validation} checks the update
kernels and the estimators against a case with an exact answer. It says nothing
about whether a given run is long enough, which in a geometry where three out
of four spins sit on the boundary cannot be taken for granted. This appendix
collects the dynamical checks; the protocol and the choice of observable are
stated in Sec.~\ref{sec:computational_protocol}.

Table~\ref{tab:tau} lists $\tau_{\mathrm{int}}$ at $L=9$ for the observables
entering our estimators. Two features matter. The correlation time grows by a
factor $3.6$ across the transition region, from $5.67(14)$ at $T=1.60$ to
$20.4(9)$ at $T=1.86$, so the demand on the protocol is largest on the
high-temperature side of $T^{\ast}$ rather than at $T_{c2}$. And below the
transition $\tau_{\mathrm{int}}$ of the signed magnetization is less than half
that of $|m|$, which is the signature of the percolating cluster discussed in
Sec.~\ref{sec:computational_protocol}.

Equilibration was checked by running, at each of four temperatures spanning the
transition region at $L=9$, one chain from the fully ordered configuration and
one from a random configuration, recording from the first step and comparing
$U_4^{\mathrm{tot}}$ measured over a fixed window of $4\times10^{4}$ steps as a
function of the number of steps discarded. The two protocols agree at
$N_{\mathrm{eq}}=2\times10^{4}$ to within $0.4$--$2.5$ standard deviations, and
neither shows a systematic drift as the discarded portion is lengthened to
$10^{5}$ steps. Incomplete equilibration is therefore excluded as a source of
bias at the largest size.

\begin{table}[t]
\centering
\caption{Integrated autocorrelation time in hybrid steps at $L=9$, for the
observables entering Eqs.~\eqref{eq:chi_region} and~\eqref{eq:Binder_def},
obtained with the automatic windowing procedure of
Ref.~\cite{MadrasSokal1988}. The last column gives the number of
autocorrelation times contained in one jackknife block of the production runs.
The signed magnetization, whose correlation time is reported for comparison
only, decorrelates faster than $|m|$ below the transition region because the
Wolff cluster percolates and reverses the global sign.}
\label{tab:tau}
\begin{ruledtabular}
\begin{tabular}{cccccc}
$T$ & $|m|$ & $m^2$ & $m^4$ & $E$ & block$/\tau_{\mathrm{int}}$ \\ \hline
1.60 &  5.67(14) &  5.60(14) &  5.21(13) & 4.66(11) & 353 \\
1.70 &  9.79(32) &  9.20(29) &  7.94(24) & 5.06(12) & 204 \\
1.78 & 15.24(63) & 14.31(57) & 11.85(43) & 4.46(10) & 131 \\
1.86 & 20.4(9)   & 20.0(9)   & 17.4(7)   & 4.08(9)  &  98 \\
\end{tabular}
\end{ruledtabular}
\end{table}

There is one prediction worth recording in advance, because it is falsifiable
and comes from a different part of the analysis. The jackknife uncertainties of
$U_4^{\mathrm{tot}}$ grow smoothly with size, from $0.0028$ at $L=4$ to
$0.0084$ at $L=9$, that is $\sigma(U_4)\sim N^{0.175}$. The Binder cumulant is
dimensionless and its intrinsic fluctuation does not grow with $N$ at a
critical point, so at fixed $N_{\mathrm{mc}}$ this growth must be attributed to
$\sigma\sim(\tau_{\mathrm{int}}/N_{\mathrm{mc}})^{1/2}$, giving
$\tau_{\mathrm{int}}\sim N^{0.35}$ and predicting
$\tau_{\mathrm{int}}\simeq10$ hybrid steps at $L=9$ near $T^{\ast}$, against
the $6.25$ measured at $L=8$. Critical slowing down is therefore present and
does grow with size, though it remains modest: at $\tau_{\mathrm{int}}\simeq10$
each jackknife block still contains some $200$ autocorrelation times.

\begin{acknowledgments}
We thank CNEA for providing computational resources. M.C.M. acknowledges
support from the Department of Mathematical Sciences at the University of
Texas at El Paso (UTEP). The \texttt{hypertiling}
library~\cite{Schrauth2024} was essential for the generation of
the hyperbolic lattices used in this work.

The analysis scripts that support the findings of this
study are available from the corresponding author upon reasonable request.
\end{acknowledgments}

\bibliography{references_v2}

\end{document}